\documentclass[aps,prl,showpacs]{revtex4}
\usepackage{amssymb}

\usepackage{amsmath}
\usepackage{graphicx}
\usepackage{bm}

\begin{document}

\title{Maximal Subgroups of the Coxeter Group $W(H_4)$ and Quaternions}
\date{\today}
\author{Mehmet Koca}
\email{kocam@squ.edu.om}
\affiliation{Department of Physics, College of Science, Sultan Qaboos University, PO Box
36, Al-Khod 123, Muscat, Sultanate of Oman}
\author{Ramazan Ko\c{c}}
\email{koc@gantep.edu.tr}
\affiliation{Department of Physics, Faculty of Engineering University of Gaziantep, 27310
Gaziantep, Turkey}
\author{Muataz Al-Barwani}
\email{muataz@squ.edu.om}
\affiliation{Department of Physics, College of Science, Sultan Qaboos University, PO Box
36, Al-Khod 123, Muscat, Sultanate of Oman}
\author{Shadia Al-Farsi}
\email{shadia@pdo.co.om}
\affiliation{Department of Physics, College of Science, Sultan Qaboos University, PO Box
36, Al-Khod 123, Muscat, Sultanate of Oman}

\begin{abstract}
The largest finite subgroup of $O(4)$ is the noncrystallographic Coxeter
group $W(H_{4})$ of order 14400. Its derived subgroup is the largest finite
subgroup $W(H_{4})/Z_{2}$ of $SO(4)$ of order 7200. Moreover, up to
conjugacy, it has five non-normal maximal subgroups of orders 144, two 240,
400 and 576. Two groups $\left[ W(H_{2})\times W(H_{2})\right] \times Z_{4}$
and $W(H_{3})\times Z_{2}$ possess noncrystallographic structures with
orders 400 and 240 respectively. The groups of orders 144, 240 and 576 are
the extensions of the Weyl groups of the root systems of $SU(3)\times SU(3)$%
, $SU(5)$ and $SO(8)$ respectively. We represent the maximal subgroups of $%
W(H_{4})$ with sets of quaternion pairs acting on the quaternionic root
systems.
\end{abstract}

\pacs{02.20.Bb}
\keywords{Group Structure, Quaternions, Subgroup structure, M-Theory.}
\maketitle

\section{Introduction}

The noncrystallographic Coxeter group $W(H_{4})$ of order 14400 generates
some interests \cite{one} for its relevance to the quasicrystallographic
structures in condensed matter physics \cite{two,three} as well as its
unique relation with the $E_{8}$ gauge symmetry associated with the
heterotic superstring theory \cite{four}. The Coxeter group $W(H_{4})$ \cite%
{five} is the maximal finite subgroup of $O(4)$, the finite subgroups of
which have been classified by du Val \cite{six} and by Conway and Smith \cite%
{seven}. It is also one of the maximal subgroups of the Weyl group $W(E_{8})$
splitting 240 nonzero roots of $E_{8}$ into two equal size disjoint sets.
One set can be represented by the icosians $q$ (quaternionic elements of the
binary icosahedral group $I$) and the remaining set is $\sigma q$ \cite%
{eight,nine} where $\sigma =\frac{1-\sqrt{5}}{2}$. Embedding of $W(H_{4})$
in $W(E_{8})$ is studied in detail in references \cite{ten} and \cite{eleven}%
.

In this paper we study the maximal subgroups of $W(H_{4})$ and show that it
possesses, up to conjugacy, five maximal subgroups of orders $144$, $240$, $%
400$ and $576$. They correspond to the symmetries of certain Dynkin and
Coxeter diagrams. Another obvious maximal subgroup $W(H_{4})^{\prime }\equiv
W(H_{4})/Z_{2}$ of $W(H_{4})$ of order $7200$ is also the largest finite
subgroup of $SO(4)$. Two groups of orders $400$ and $240$ are related to the
symmetries of the noncrystallographic Coxeter graphs $H_{2}\oplus
H_{2}^{\prime }$ and $H_{3}$ respectively. The remaining groups of orders $%
144$, $240$ and $576$ are associated with the symmetries of the
crystallographic root systems $A_{2}\oplus A_{2}^{\prime }$, $A_{4}$ and $%
D_{4}$ respectively (we interchangeably use $A_{2}\approx SU(3),A_{4}\approx
SU(5),D_{4}\approx SO(8)$). Embeddings of these groups in $W(H_{4})$ are not
trivial and the main objective of this work is to clarify this issue.

In the context of quaternionic representation of the $H_{4}$ root system by
the elements of binary icosahedral group, identifications of the
quaternionic root systems of the maximal subgroups will be simple. The
maximal subgroups of orders $144$, $400$ and $576$ associated with the root
systems $A_{2}\oplus A_{2}^{\prime },H_{2}\oplus H_{2}^{\prime }$ and $D_{4}$
respectively have close relations with the maximal subgroups of the binary
icosahedral group $I$ since it has three maximal subgroups. However the
maximal subgroups associated with the Coxeter diagram $H_{3}$ and the Dynkin
diagram $A_{4}$ do not have such correspondences and some care should be
undertaken.

In Sec.~\ref{sec:two} we introduce the root system of $H_{4}$ in terms of
the icosions $I$ and identify the maximal subgroups of the binary
icosahedral group. We discuss briefly the method as to how the group
elements of $W(H_{4})$ are obtained from icosians. In Sec.~\ref{sec:three},
we study in terms of quaternions, the group structure associated with the
graph $A_{2}\oplus A_{2}^{\prime }$. Sec.~\ref{sec:four} is devoted to a
similar analysis of the symmetries of the noncrystallographic root system $%
H_{2}\oplus H_{2}^{\prime }$. In Sec.~\ref{sec:five} we deal with the
extension of $W(D_{4})$ with a cyclic symmetry of its Dynkin diagram, the
group of order $576$, which was also studied in a paper of ours \cite{twelve}
in a different context. Sec.~\ref{sec:six} is devoted to the study of the
automorphism of the $H_{3}$ root system and we show that it is, up to
conjugacy, a group of order 240 when it acts in the 4-dimensional space. In
Sec.~\ref{sec:seven} we study the somewhat intricate structure of the
automorphism group of $A_{4}$ using quaternions.

\section{$H_4$ with Icosians}

\label{sec:two}

The root system of the Coxeter diagram $H_{4}$ consists of $120$ roots which
can be represented by the quaternionic elements of the binary icosahedral
group $I$ given in Table~\ref{tab:one} (so is called the group of icosians).
They can be generated by reflections on the simple roots depicted in Fig~\ref%
{fig:one}.

Any real quaternion can be written as $%
q=q_{0}+q_{1}e_{1}+q_{2}e_{2}+q_{3}e_{3}$ where $q_{a}$ $(a=0,1,2,3)$ are
real numbers and pure quaternion units\footnote{%
Any unit quaternion $q=-\overline{q}$ is a pure quaternion, where the
quaternion conjugate is $\overline{q}=q_{0}-q_{1}e_{1}-q_{2}e_{2}-q_{3}e_{3}$%
} $e_{i}$ ($i=1,2,3$) satisfy the well-known relations
\begin{equation}
e_{i}e_{j}=-\delta _{ij}+\epsilon _{ijk}e_{k}\quad (i,j,k=1,2,3)
\end{equation}%
where $\epsilon _{ijk}$ is the Levi-Civita symbol. The scalar product of two
quaternions $p$ and $q$ is defined by
\begin{equation}
(p,q)=\frac{1}{2}(p\overline{q}+q\overline{p})
\end{equation}%
which leads to the norm $N(q)=(q,q)=q\overline{q}%
=q_{0}^{2}+q_{1}^{2}+q_{2}^{2}+q_{3}^{2}$.
\begin{table}[h]
\caption{Conjugacy classes of the binary icosahedral group $I$ represented
by quaternions.(Here $\protect\tau =\frac{1+\protect\sqrt{5}}{2}$ and $%
\protect\sigma =\frac{1-\protect\sqrt{5}}{2}$)}
\label{tab:one}%
\begin{tabular}{cl}
&  \\ \hline
Conjugacy Classes & \multicolumn{1}{|l}{Elements of the conjugacy classes
also denoted by their numbers} \\
and orders of elements & \multicolumn{1}{|l}{(Cyclic permutations in $%
e_{1},e_{2},e_{3}$ should be added if not included)} \\ \hline
$1$ & \multicolumn{1}{|l}{$1$} \\
$2$ & \multicolumn{1}{|l}{$-1$} \\
$10$ & \multicolumn{1}{|l}{$12_{+}:\frac{1}{2}(\tau \pm e_{1}\pm \sigma
e_{3})$} \\
$5$ & \multicolumn{1}{|l}{$12_{-}:\frac{1}{2}(-\tau \pm e_{1}\pm \sigma
e_{3})$} \\
$10$ & \multicolumn{1}{|l}{$12_{+}^{\prime }:\frac{1}{2}(\sigma \pm e_{1}\pm
\tau e_{2})$} \\
$5$ & \multicolumn{1}{|l}{$12_{-}^{\prime }:\frac{1}{2}(-\sigma \pm e_{1}\pm
\tau e_{2})$} \\
$6$ & \multicolumn{1}{|l}{$20_{+}:\frac{1}{2}(1\pm e_{1}\pm e_{2}\pm e_{3}),%
\frac{1}{2}(1\pm \tau e_{1}\pm \sigma e_{2})$} \\
$3$ & \multicolumn{1}{|l}{$20_{-}:\frac{1}{2}(-1\pm e_{1}\pm e_{2}\pm e_{3}),%
\frac{1}{2}(-1\pm \tau e_{1}\pm \sigma e_{2})$} \\
$4$ & \multicolumn{1}{|l}{$30:%
\begin{array}{l}
15_{+}:e_{1},e_{2},e_{3},\frac{1}{2}(\sigma e_{1}\pm \tau e_{2}\pm e_{3}) \\
15_{-}:-e_{1},-e_{2},-e_{3},\frac{1}{2}(-\sigma e_{1}\pm \tau e_{2}\pm e_{3})%
\end{array}%
$} \\ \hline
\end{tabular}%
\end{table}

\begin{figure}[h]
\begin{center}
\begin{picture}(300,70)(0,0)
\put(25,30){$-e_1$}
\put(40,15){\circle*{5}}
\put(40,15){\line(60,0){60}} \put(65,30){5}
\put(100,15){\circle*{5}}  \put(60,0){$\frac{1}{2}(\tau e_1+e_2+\sigma e_3)$}
\put(100,15){\line(60,0){60}}
\put(160,15){\circle*{5}}  \put(150,30){$-e_2$}
\put(160,15){\line(60,0){60}}
\put(220,15){\circle*{5}}  \put(230,15){$\frac{1}{2}(\sigma+e_2+\tau e_3)$}
\end{picture}
\end{center}
\caption{The Coxeter diagram of $H_{4}$ with quaternionic simple roots}
\label{fig:one}
\end{figure}
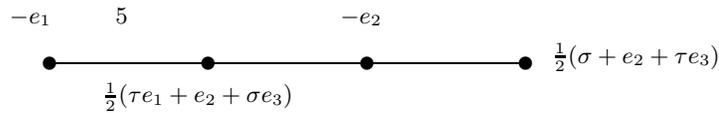

The icosians of the binary icosahedral group $I$ are classified in Table~\ref%
{tab:one} according to the conjugacy classes. Denote by $p$, $q$ and $r$ any
three elements of $I$. The transformations defined by
\begin{eqnarray}
\left[ p,r\right] &:&q\rightarrow pqr  \label{eq:3a} \\
\left[ p,r\right] ^{\ast } &:&q\rightarrow p\overline{q}r  \label{eq:3b}
\end{eqnarray}%
preserve the norm $q\overline{q}=\overline{q}q$ as well as leave the set of
icosians $I$ intact so that the pairs $[p,r]$ and $[p,r]^{\ast }$ represent
the group elements of $W(H_{4})$. Since $[p,r]=[-p,-r]$ and $[p,-r]=[-p,r]$
the $W(H_{4})$ consists of $120\times 120=14400$ elements. Details can be
found in reference~\cite{eleven}. The element $[1,1]^{\ast }$ acts as a
conjugation, $[1,1]^{\ast }:q\rightarrow \overline{q}$, which is the
normalizer of the subgroup $SO(4)$ so that $O(4)$ can be written as the
semi-direct product $O(4)\approx SO(4)\times Z_{2}$. It follows from this
structure that one of the maximal subgroup $W(H_{4})^{\prime }$ of order $%
7200$ is obviously a maximal finite subgroup of $SO(4)$. The group $%
W(H_{4})^{\prime }$ can be represented by the pair $[p,r]$, $p,r\in I$,
possessing $42$ conjugacy classes. It is easily found from the structure of
the conjugacy classes of $I$ in Table~\ref{tab:one} that the maximal
subgroups of $I$ can be generated by the sets of quaternions\footnote{%
For a detailed study see reference~\cite{eleven}}

\begin{eqnarray}
\mathrm{Dihedral\ group\ of\ order\ 12} &&\quad :\quad \frac{1}{2}(1+\tau
e_{1}+\sigma e_{2}),e_{3}  \label{eq:5a} \\
\mathrm{Dihedral\ group\ of\ order\ 20} &&\quad :\quad \frac{1}{2}(\tau
+\sigma e_{1}+e_{2}),e_{3}  \label{eq:5b} \\
\mathrm{Binary\ tetrahedral\ group\ of\ order\ 24} &&\quad :\quad \frac{1}{2}%
(1+e_{1}+e_{2}+e_{3}),e_{3}  \label{eq:5c}
\end{eqnarray}%
These generators are the representative elements of the conjugacy classes of
the maximal subgroups of $I$. The binary icosahedral group has two
quaternionic irreducible representations. If we denote by $I^{\prime }$ the
other quaternionic representation it can be obtained from $I$ by
interchanging $\sigma \leftrightarrow \tau $ in Table~\ref{tab:one}. We will
use both representations in section~\ref{sec:seven}.

\section{The maximal subgroup of order 144}

\label{sec:three}

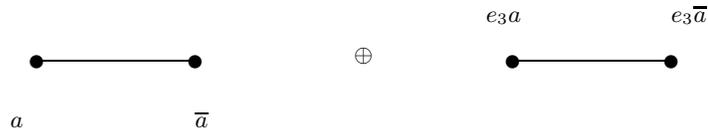
\begin{figure}[h]
\begin{center}
\begin{picture}(260,70)(0,0)
\put(10,0){$a$}
\put(20,25){\circle*{5}}
\put(20,25){\line(60,0){60}}
\put(80,25){\circle*{5}}  \put(80,0){$\overline{a}$}
\put(140,25){$\oplus$}
\put(200,25){\circle*{5}}  \put(190,40){$e_3a$}
\put(200,25){\line(60,0){60}}
\put(260,25){\circle*{5}}  \put(260,40){$e_3\overline{a}$}
\end{picture}
\end{center}
\caption{Dynkin diagram of $A_{2}\oplus A_{2}^{\prime }$with the elements of
dihedral group of order $12$\ ($a=\frac{1}{2}(1+\protect\tau e_{1}+\protect%
\sigma e_{2})$ ,\quad $\overline{a}=\frac{1}{2}(1-\protect\tau e_{1}-\protect%
\sigma e_{2})$ ).}
\label{fig:two}
\end{figure}

A little exercise shows that the set of elements in Eq.~\ref{eq:5a}
constitute the root system of the Lie algebra $A_{2}\oplus A_{2}^{\prime }$
where the simple roots are given in Figure~\ref{fig:two}.

We have used one of the four dimensional irreducible matrix representations
of $W(H_{4})$ to identify its maximal subgroups via computer calculations
which resulted in those six maximal subgroups. Our main objective here is to
construct the elements of the maximal subgroups using pairs of quaternions
described in (\ref{eq:3a}-\ref{eq:3b}). In each section we give arguments -
though not very rigorous - that the groups of concern are maximal in $%
W(H_{4})$. We take the computer calculations as evidence for the
completeness of the maximal subgroups. The rigorous proof is beyond the
scope of this paper.

It follows that the sets ($\pm a,\pm \overline{a},\pm 1$), ($a^{6}=1$) and $%
(\pm e_{3}a,\pm e_{3}\overline{a},\pm e_{3})$ constitute the roots of $A_{2}$
and $A_{2}^{\prime }$ respectively. Then the generators of the Weyl groups $%
W(A_{2})$ and $W(A_{2}^{\prime })$ are the group elements\footnote{%
The reflection of a quaternion $q$ in the plane orthogonal to a root, say a
unit quaternion $a$, can be written as $r_{a}(q)=q-2(q,a)a=-a\overline{q}a$.}
\begin{eqnarray}
r_{1} &=&[a,-a]^{\ast },\quad r_{2}=[\overline{a},-\overline{a}]^{\ast }
\label{eq:6a} \\
r_{1}^{\prime } &=&[e_{3}a,-e_{3}a]^{\ast },\quad r_{2}^{\prime }=[e_{3}%
\overline{a},-e_{3}\overline{a}]^{\ast }.  \label{eq:6b}
\end{eqnarray}%
The sets of elements of $W(A_{2})$ and $W(A_{2}^{\prime })$ can be easily
generated by (\ref{eq:6a}) and (\ref{eq:6b}):
\begin{eqnarray}
&&W(A_{2})\approx D_{3}\approx S_{3}:[a,-a]^{\ast },[\overline{a},-\overline{%
a}]^{\ast },[1,-1]^{\ast },[a,a],[\overline{a},\overline{a}],[1,1]
\label{7a} \\
&&W(A_{2}^{\prime })\approx D_{3}\approx S_{3}:  \notag \\
&&[e_{3}a,-e_{3}a]^{\ast },[e_{3}\overline{a},-e_{3}\overline{a}]^{\ast
},[e_{3},-e_{3}]^{\ast },[\overline{a},a],[a,\overline{a}],[1,1].  \label{7b}
\end{eqnarray}%
Note that the longest elements of the respective groups are $%
w_{0}=[-1,1]^{\ast }$ and $w_{0}^{\prime }=[e_{3},-e_{3}]^{\ast }.$ The
automorphism groups $Aut(A_{2})$ and $Aut(A_{2}^{\prime })$ of the
respective root systems are the extensions of the Weyl groups by the
respective Dynkin diagram symmetries. The diagram symmetries of $A_{2}$ and $%
A_{2}^{\prime }$ are obtained by the exchange of simple roots $%
a\leftrightarrow \overline{a}$ and $e_{3}a\leftrightarrow e_{3}\overline{a}$
which respectively lead to the groups
\begin{equation}
Aut(A_{2})\approx W(A_{2})\rtimes \gamma ,\quad Aut(A_{2}^{\prime })\approx
W(A_{2})^{\prime }\rtimes \gamma ^{\prime }.  \label{eq:8}
\end{equation}%
where $\gamma =[1,1]^{\ast }$ and $\gamma ^{\prime }=[e_{3},e_{3}]^{\ast }$.
The groups in (\ref{eq:8}) having equal orders $12$ commute with each other.
It is interesting to observe that the automorphism group $Aut(A_{2})$ can be
written as a sum of two cosets in two different ways,
\begin{equation}
Aut(A_{2})=\{W(A_{2}),\gamma W(A_{2})\}  \label{eq:8a}
\end{equation}%
and
\begin{equation}
Aut(A_{2})=\{W(A_{2}),w_{0}\gamma W(A_{2})\}  \label{eq:8b}
\end{equation}%
where $w_{0}\gamma =c=[-1,1],\ c^{2}=[1,1]$ which commutes with $W(A_{2})$.
Therefore the group $Aut(A_{2})$ can also be written as $Aut(A_{2})\approx
W(A_{2})\times Z_{2}$, where $Z_{2}$ is generated by the element $c$. A
similar analysis is true for the diagram $A_{2}^{\prime }$ where $%
w_{0}^{\prime }\gamma ^{\prime }=c$. The direct product of the two Weyl
groups and the group $Z_{2}$ is of order $72$. The elements of the cyclic
group $Z_{4}=<[e_{3},1]>=\{[\pm 1,1],[\pm e_{3},1]\}$ transform the two
groups $W(A_{2})$ and $W(A_{2}^{\prime })$ to each other by conjugation
where $[e_{3},1]^{2}=c$. Since the group $W(A_{2})\times W(A_{2}^{\prime })$
is invariant under $Z_{4}$ by conjugation and the only common element to
both groups is the unit element $[1,1]$ then one can extend $W(A_{2})\times
W(A_{2}^{\prime })$ by $Z_{4}$ to $[W(A_{2})\times W(A_{2})]\rtimes Z_{4}$.
The group $[W(A_{2})\times W(A_{2}^{\prime })]\rtimes Z_{4},$ of four cosets
$[1,1][W(A_{2})\times W(A_{2}^{\prime })],[-1,1][W(A_{2})\times
W(A_{2}^{\prime })],[e_{3},1][W(A_{2})\times W(A_{2}^{\prime })]$, and $%
[-e_{3},1][W(A_{2})\times W(A_{2}^{\prime })]$ has order $144.$ Moreover it
contains $W(A_{2})\times W(A_{2}^{\prime })\times Z_{2}$ as a maximal
subgroup of index 2. The element $[e_{3},1]$ acts by exchanging the
reflections $r_{1}$and $r_{2}^{\prime }$ and $r_{2}$ and $r_{1}^{\prime }.$
This is a diagram automorphism of the root system of $A_{2}\oplus
A_{2}^{\prime }$. Since the Weyl group is extended by the cyclic group $%
Z_{4} $ generated by the element $[e_{3},1]$ the extended group is nothing
but $[W(A_{2})\times W(A_{2}^{\prime })]\rtimes Z_{4}\approx Aut[A_{2}\oplus
A_{2}^{\prime }]$

In terms of the quaternion pairs in (\ref{eq:3a}, \ref{eq:3b}) the above
group can be written compactly
\begin{equation}
\lbrack p,r],[p,r]^{\ast }\quad \mathrm{with}\quad p,r\in \{\pm a,\pm
\overline{a},\pm 1,\pm e_{3}a,\pm e_{3}\overline{a},\pm e_{3}\}.
\label{eq:10}
\end{equation}

To prove that the group $[W(A_{2})\times W(A_{2}^{\prime })]\rtimes Z_{4}$
is maximal in $W(H_{4})$ we argue as follows. Let us denote by $[b,1]$ an
element of $W(H_{4})$ not belonging to $[W(A_{2})\times W(A_{2}^{\prime
})]\rtimes Z_{4}$, i.e. $b$ is not one of those quaternions in the set of
roots $A_{2}\oplus A_{2}^{\prime }$ in (\ref{eq:10}). The products of $[b,1]$
with the group elements $[p,r]$ and $[p,r]^{\ast }$ of (\ref{eq:10}) would
yield the elements $[pb,r],[bp,r],[bp,r]^{\ast }$ and $[p,\overline{b}%
r]^{\ast }$. These elements obviously do not belong to the group $%
[W(A_{2})\times W(A_{2})]\rtimes Z_{4}$. The elements $pb$ and $\overline{b}%
r $ and those elements obtained by repetitive multiplications will generate
the binary icosahedral group $I$ not any subgroups of it, since the dihedral
group of order $12$ is already a maximal subgroup of $I$. Therefore the
group generated in this manner will be the whole group $W(H_{4})=%
\{[I,I],[I,I]^{\ast }\}$ and the group $[W(A_{2})\times W(A_{2}^{\prime
})]\rtimes Z_{4}$ is, up to conjugacy, a maximal subgroup of $W(H_{4})$.

Similar arguments apply for the maximality of the groups $[W(H_{2})\times
W(H_{2}^{\prime })]\rtimes Z_{4}$ and $W(D_{4})\rtimes Z_{3}$ as they are
generated from the maximal dihedral subgroup of order $20$ and the maximal
binary tetrahedral subgroup of the binary icosahedral group $I$.

\section{The symmetry group of $H_2\oplus H_2^\prime$ of order 400}

\label{sec:four}

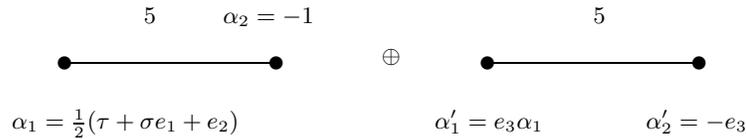
\begin{figure}[h]
\begin{center}
\begin{picture}(260,70)(0,0)
\put(0,0){$\alpha_1=\frac{1}{2}(\tau +\sigma e_1 + e_2)$}
\put(20,25){\circle*{5}}
\put(20,25){\line(80,0){80}} \put(50,40){5}
\put(100,25){\circle*{5}}  \put(80,40){$\alpha_2= -1$}
\put(140,25){$\oplus$}
\put(180,25){\circle*{5}}  \put(160,0){$\alpha_1^\prime = e_3\alpha_1$}
\put(180,25){\line(80,0){80}} \put(220,40){5}
\put(260,25){\circle*{5}}  \put(240,0){$\alpha_2^\prime = -e_3$}
\end{picture}
\end{center}
\caption{Coxeter diagram of $H_{2}\oplus H_{2}^{\prime }$ \ with
quaternions. }
\label{fig:three}
\end{figure}

A similar analysis to the one studied in Sec.\ref{sec:three} can be pursued
by introducing the root system of the noncrystallographic Coxeter diagram $%
H_{2}\oplus H_{2}^{\prime }$. The Coxeter diagram of the system with
quaternions is shown in Figure~\ref{fig:three}.
If we denote by $b=\frac{1}{2}(\tau +\sigma e_{1}+e_{2}),$ then $b^{5}=-1$
and the set of roots of $H_{2}\oplus H_{2}^{\prime }$ will be given by
\begin{eqnarray}
H_{2} &:&\pm b,\pm b^{2},\pm \overline{b}^{2},\pm \overline{b},\pm 1
\label{eq:11a} \\
H_{2}^{\prime } &:&\pm e_{3}b,\pm e_{3}b^{2},\pm e_{3}\overline{b}^{2},\pm
e_{3}\overline{b},\pm e_{3}.  \label{eq:11b}
\end{eqnarray}%
We note that (\ref{eq:11a}-\ref{eq:11b}) are the quaternion elements of the
dihedral group of order $20$ given in (\ref{eq:5b}). They can be more
compactly represented as $H_{2}:\pm b^{m},H_{2}^{\prime }:\pm e_{3}b^{m}$ ($%
m=1,2,3,4,5$). The group $W(H_{2})$ is generated by the reflections in the
hyperplanes orthogonal to the simple roots $\alpha _{1}$ and $\alpha _{2}.$ $%
W(H_{2})$ can be enumarated explicitly and compactly:
\begin{equation}
W(H_{2})=\left\{ \left[ b^{m},-b^{m}\right] ^{\ast }\right\} \cup \left\{ %
\left[ b^{m},b^{m}\right] \right\} ,\quad m=1,\ldots ,5.  \label{eq12}
\end{equation}%
Similarly, we can obtain the group elements of $W(H_{2}^{\prime })$ as
\begin{equation}
\lbrack e_{3}b^{m},-e_{3}b^{m}]^{\ast },\quad \lbrack b^{m},\overline{b}%
^{m}],\quad (m=1,2,3,4,5).
\end{equation}%
Now each group can be extended to $Aut(H_{2})$ and $Aut(H_{2}^{\prime })$ by
the respective diagram symmetries $\delta _{2}:\alpha _{1}\leftrightarrow
\alpha _{2}$ and $\delta _{2}^{\prime }:\alpha _{1}^{\prime }\leftrightarrow
\alpha _{2}^{\prime }$ where $\delta _{2}=[\overline{b}^{2},\overline{b}%
^{2}]^{\ast }$ and $\delta _{2}^{\prime }=[e_{3}\overline{b}^{2},e_{3}%
\overline{b}^{2}]^{\ast }$ . Each automorphism group possesses $20$ elements
and can be represented by the following sets of elements:
\begin{eqnarray}
Aut(H_{2}) &:&[b^{m},\pm b^{m}],[b^{m},\pm b^{m}]^{\ast }  \label{eq:14a} \\
Aut(H_{2}^{\prime }) &:&[e_{3}b^{m},\pm e_{3}b^{m}]^{\ast },[b^{m},\pm
\overline{b}^{m}],\quad m=1,\ldots ,5.  \label{eq:14b}
\end{eqnarray}%
An analysis similar to (\ref{eq:8a}) and (\ref{eq:8b}) can be carried out.
We note that the longest elements of $W(H_{2})$ and $W(H_{2}^{\prime })$ are
$w_{0}=r_{1}r_{2}r_{1}r_{2}r_{1}=[-b^{3},b^{3}]^{\ast }$, where $r_{1}$ and $%
r_{2}$ are the reflection generators of $W(H_{2})$ on the roots $\alpha _{1}$
and $\alpha _{2}$ respectively. A similar consideration leads to the longest
element of $W(A_{2}^{\prime })$ $w_{0}^{\prime
}=[-e_{3}b^{3},e_{3}b^{3}]^{\ast }$. We can obtain the products $w_{0}\delta
=w_{0}^{\prime }\delta ^{\prime }=[-1,1]=c$ which commutes with the elements
of $W(H_{2})$ and $W(H_{2}^{\prime })$. Therefore we can write the groups $%
Aut(H_{2})$ and $Aut(H_{2}^{\prime })$ as $Aut(H_{2})\approx W(H_{2})\times
Z_{2}$ and $Aut(H_{2}^{\prime })\approx W(H_{2}^{\prime })\times Z_{2}$. One
can also check that $W(H_{2})$ and $W(H_{2}^{\prime })$ are conjugates under
the action of $Z_{4}=<[e_{3},1]>$ and thus the maximal group of order $400$
has the structure $[W(H_{2})\times (H_{2}^{\prime })]\rtimes Z_{4}$. The
group elements can be put into the form $[p,q],[p,q]^{\ast }$ where $p$ and $%
q$ take values from the set of roots of $H_{2}\oplus H_{2}^{\prime }$
\begin{equation}
p,q\in \{\pm b^{m},\pm e_{3}b^{m}\},\quad m=1,2,3,4,5.
\end{equation}%
The group generated by $p,q$ pairs consist of $400$ elements. The group $%
Aut[H_{2}\oplus H_{2}^{\prime }]\approx \lbrack W(H_{2})\times
W(H_{2}^{\prime })]\rtimes Z_{4}$ is maximal as a consequence of the
arguments used in Sec.~\ref{sec:three}.

\section{The maximal subgroup $W(D_4)\rtimes Z_3$ of order 576}

\label{sec:five}

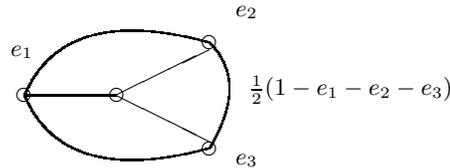
\begin{figure}[th]
\begin{picture}(200,70)(0,0)
\put(60,30){$e_1$}
\put(65,15) {\circle{5}}
\put(65,15) {\line(35,0){35}}
\put(100,15) {\circle{5}}
\put(150,15) {$\frac{1}{2}(1-e_1-e_2-e_3)$}
\put(100,15) {\line (2,1){35}}
\put(135,35) {\circle{5}}
\put(145,45) {$e_2$}
\put(100,15){\line(2,-1){35}}
\put(135,-5){\circle{5}}  \put(145,-11){$e_3$}
\qbezier(65,15)(80,50)(135,35)
\qbezier(65,15)(80,-20)(135,-5)
\qbezier(135,35)(150,20)(135,-5)
\end{picture}
\caption{Dynkin diagram of $SO(8)$ \ with quaternionic simple roots.}
\label{fig:four}
\end{figure}

This group has been discussed in reference~\cite{twelve} in a different
context. A brief consideration may be in order. The root system of $%
D_{4}\approx SO(8)$ can be generated by the simple roots of quaternions
shown in Figure~\ref{fig:four}.

The set of roots constitute the elements of the binary tetrahedral group $T$%
, a maximal subgroup of the binary icosahedral group $I$. The Weyl group $%
W(D_{4})$ of order $192$ with $13$ conjugacy classes can be generated by the
reflection generators $[-e_{1},e_{1}]^{\ast },[-e_{2},e_{2}]^{\ast
},[-e_{3},e_{3}]^{\ast }$ and $[\frac{1}{2}(-1+e_{1}+e_{2}+e_{3}),\frac{1}{2}%
(1-e_{1}-e_{2}-e_{3})^{\ast }]$. The automorphism group of the $SO(8)$ root
system with the inclusion of $Z_{3}$-Dynkin diagram symmetry is generated by
$W(D_{4})$ along with $[\frac{1}{2}(1+e_{1}+e_{2}+e_{3}),\frac{1}{2}%
(1-e_{1}-e_{2}-e_{3})],$ a cyclic symmetry of the diagram for $W(D_{4}).$
This gives a subgroup of $W(H_{4})$ of order $576$. We note in passing that
the full automorphism group of the root system of $SO(8)$ is $%
W(D_{4})\rtimes S_{3}$ of order $1152$ and it is not a subgroup of $W(H_{4})$%
.

When the quaternions $p,q$ take values from the root system of $SO(8)$
\begin{equation}
p,q\in T=\{\pm 1,\pm e_{1},\pm e_{2},\pm e_{3},\frac{1}{2}\left( \pm 1\pm
e_{1}\pm e_{2}\pm e_{3}\right) \}
\end{equation}%
the group consisting of elements $[p,q],[p,q]^{\ast }$ forms the desired
maximal subgroup of order 576. Under the action of $W(D_{4})$ the 120 roots
of $H_{4}$ split into four sets of elements
\begin{equation}
120=24+32_{1}+32_{2}+32_{3}.
\end{equation}%
The first $24$ represents the roots of $SO(8)$ and each set of $32$ elements
is associated with one of the quaternionic units $e_{i}.$ The $Z_{3}$
symmetry of the Dynkin diagram permutes these three sets of $32$ elements.
One of these three sets of $32$ elements can be represented by the
quaternions
\begin{equation}
32_{1}:\frac{1}{2}(\pm \tau \pm e_{1}\pm \sigma e_{3}),\frac{1}{2}(\pm
\sigma \pm e_{1}\pm \tau e_{2}),\frac{1}{2}(\pm 1\pm \tau e_{1}\pm \sigma
e_{2}),\frac{1}{2}(\pm \sigma e_{1}\pm \tau e_{2}\pm e_{3}).  \label{eq:18}
\end{equation}%
The other sets of $32$ elements are obtained by cyclic permutations of $%
e_{1},e_{2}$ and $e_{3}$. However, under the action of the group $%
W(D_{4})\rtimes Z_{3}$ the $120$ elements split as $120=24+96$. The group $%
W(D_{4})\rtimes Z_{3}$ is maximal up to conjugacy in the group $W(H_{4})$ as
argued in Sec.~\ref{sec:three}.

\section{The maximal subgroup $W(H_3)\times Z_2$}

\label{sec:six}

\begin{figure}[th]
\begin{center}
\begin{picture}(200,70)(0,0)
\put(25,30){$-e_1$}
\put(40,15){\circle*{5}}
\put(40,15){\line(60,0){60}} \put(65,30){5}
\put(100,15){\circle*{5}}  \put(60,0){$\frac{1}{2}(\tau e_1+e_2+\sigma e_3)$}
\put(100,15){\line(60,0){60}}
\put(160,15){\circle*{5}}  \put(150,30){$-e_2$}
\end{picture}
\end{center}
\caption{The Coxeter diagram of $H_{3}$ with pure quaternions of $I$.}
\label{fig:five}
\end{figure}
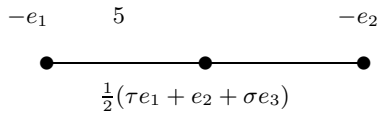

The Coxeter group $W(H_{3})$ has many applications in physics. It is the
symmetry of an icosahedron with inversion and is isomorphic to the group $%
A_{5}\times Z_{2}$ of order $120$ where $A_{5}$ is the group of even
permutations of five letters. The $C_{60}$ molecule is the popular example
possessing a truncated icosahedral structure. Some metal alloys also display
quasicrystallographic aspects with $H_{3}$ symmetry. The Coxeter diagram of $%
H_{3}$ can be represented by pure quaternionic simple roots as shown in
Figure~\ref{fig:five}. The roots of $H_{3}$ obtained by reflections in the
hyperplanes orthogonal to the simple roots of $H_{3}$~\cite{thirteen} form a
set of $30$ pure quaternions which constitute one conjugacy class in $I$
(see Table~\ref{tab:one}). Deleting the right most root $\frac{1}{2}(\sigma
+e_{2}+\tau e_{3})$ of $H_{4}$ in Figure~\ref{fig:one} we obtain the root
diagram of $H_{3}$ as shown in Figure~\ref{fig:five}.
The generators of $W(H_{3})\quad \lbrack -e_{1},e_{1}]^{\ast },[-\frac{1}{2}%
(\tau e_{1}+e_{2}+\sigma e_{3}),\frac{1}{2}(\tau e_{1}+e_{2}+\sigma
e_{3})]^{\ast }$ and $[-e_{2},e_{2}]^{\ast }$ generate the group $W(H_{3})$
which can be put into the form
\begin{equation}
W(H_{3}):\ [p,\overline{p}],[p,\overline{p}]^{\ast }  \label{eq:19}
\end{equation}%
where $p$ takes any one of the $120$ elements of $I$. In fact $\left\{ [p,%
\overline{p}]\quad |\quad p\in I\right\} $ is isomorphic to $A_{5},$ a group
of order $60$ which is the largest finite subgroup of $SO(3)$. The $[p,%
\overline{p}]^{\ast }$ are obtained simply by multiplying $[p,\overline{p}]$
by the conjugation element $[1,1]^{\ast }$ which takes $q\rightarrow
\overline{q}$. The group element $[1,1]^{\ast }$ commutes with the elements $%
[p,\overline{p}]$ implying the structure of $W(H_{3})\approx A_{5}\times
Z_{2}$. The five conjugacy classes of $A_{5}$ consist of the sets of
elements
\begin{equation}
\lbrack 1,1],[12_{+},\overline{12}_{+}],[12_{+}^{\prime },\overline{%
12^{\prime }}_{+}],[20_{+},\overline{20}_{+}],[15_{+},\overline{15}_{+}].
\label{eq:20}
\end{equation}%
The group $W(H_{3})$ with $10$ conjugacy classes splits the roots of $H_{4}$%
, as expected, into nine sets of elements as shown in Table~\ref{tab:one},
because $\overline{p}=p^{-1}$ and thus the elements $[p,\overline{p}]$ of $%
W(H_{3})$ act by conjugation on $I$. It is obvious that the roots of $H_{3}$
are left invariant by the element $[-1,1]$ which simply takes any root of $%
H_{3}$ to its negative. But $[-1,1]$ cannot be generated from the $H_{3}$
diagram by reflections. Moreover $[-1,1]$ commutes with the elements in (\ref%
{eq:19}). Therefore the extended symmetry of the $H_{3}$ root system is $%
W(H_{3})\times Z_{2}$ for $[-1,1]^{2}=[1,1]$ . The group $W(H_{3})\times
Z_{2}$ of order 240 is maximal in $W(H_{4})$ and can be written as $%
A_{5}\times Z_{2}^{2}$. The element $[-1,1]$ transforms the negative $(-)$
and positive $(+)$ conjugacy classes of $I$ into each other $%
1\leftrightarrow -1,12_{+}\leftrightarrow 12_{-},12_{+}^{\prime
}\leftrightarrow 12_{-}^{\prime },20_{+}\leftrightarrow
20_{-},30\leftrightarrow 30$ so that the roots of $H_{4}$ decompose into
five disjoint sets consisting of elements $2,24,24^{\prime },40$ and $30$.
The second $Z_{2}$ generator $[-1,1]$ can be taken as a coset representative
and the group $W(H_{3})\times Z_{2}$ consisting of 240 elements can be
written as
\begin{equation}
\lbrack \pm p,\overline{p}],[\pm p,\overline{p}]^{\ast }.  \label{eq:21}
\end{equation}%
This is just one way of embedding $W(H_{3})\times Z_{2}$ in $W(H_{4})$.
Since the index of $W(H_{3})\times Z_{2}$ in $W(H_{4})$ is 60 there are 60
different choices for the representations of $H_{3}\times Z_{2}$ in $%
W(H_{4}) $.

\section{The $Aut(A_{4})$ as the maximal subgroup of $W(H_{4})$}

\label{sec:seven}

The automorphism group of the root system of $A_4\approx SU(5)$ is the
semi-direct product of the Weyl group $W(A_4)$ with the $Z_2$ symmetry of
the Dynkin diagram
\begin{equation}
Aut(A_4)\approx W(A_4)\rtimes Z_2.  \label{eq:23}
\end{equation}
Since the Weyl group $W(A_4)$ is isomorphic to $S_5$ of order 120 with seven
conjugacy classes, the $Aut(A_4)$ is a group of order 240. In what follows,
we will study some details of this group.

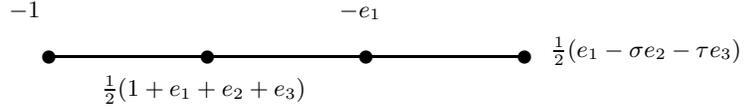
\begin{figure}[h]
\begin{picture}(200,70)(0,0)
\put(25,30){$-1$}
\put(40,15){\circle*{5}}
\put(40,15){\line(60,0){60}}
\put(100,15){\circle*{5}}  \put(60,0){$\frac{1}{2}(1+ e_1+e_2+e_3)$}
\put(100,15){\line(60,0){60}}
\put(160,15){\circle*{5}}  \put(150,30){$-e_1$}
\put(160,15){\line(60,0){60}}
\put(220,15){\circle*{5}}  \put(230,15){$\frac{1}{2}(e_1-\sigma e_2-\tau e_3)$}
\end{picture}
\caption{The Dynkin diagram of $SU(5)$ \ with quaternion simple roots.}
\label{fig:six}
\end{figure}

The Dynkin diagram of $SU(5)$ with quaternionic simple roots can be obtained
from Figure~\ref{fig:one} by deleting the left-most root ($-e_{1}$) and
adding another root to the right of the right-most root $\frac{1}{2}(\sigma
+e_{2}+\tau e_{3})$. However, for the simplicity of calculations of group
elements, we choose the root system of the Dynkin diagram $A_{4}$ as shown
in Figure~\ref{fig:six}.
The Weyl group $W(A_{4})$ is generated by reflections on the simple roots
and can be written as
\begin{eqnarray}
&&\left[ -1,1\right] ^{\ast },[-\frac{1}{2}(1+e_{1}+e_{2}+e_{3}),\frac{1}{2}%
(1+e_{1}+e_{2}+e_{3})]^{\ast },  \label{eq:24} \\
&&[-e_{1},e_{1}]^{\ast },[-\frac{1}{2}(e_{1}-\sigma e_{2}-\tau e_{3}),\frac{1%
}{2}(e_{1}-\sigma e_{2}-\tau e_{3})]^{\ast }.  \notag
\end{eqnarray}%
These elements generate a group of order 120 isomorphic to the permutation
group $S_{5}$. The group elements of $W(A_{4})$ can be put into the form
\begin{equation}
\lbrack p,-a\overline{p}^{\prime }a],[p,a\overline{p}^{\prime }a]^{\ast }.
\label{eq:25}
\end{equation}%
Here $a=\frac{1}{\sqrt{2}}(e_{2}-e_{3})$ is a pure quaternion and $p^{\prime
}$, obtained from $p$ by exchanging $\tau \leftrightarrow \sigma $, is an
element of the second quaternionic representation $I^{\prime }$ of the
binary icosahedral group obtained from $I$ by exchanging $\tau
\leftrightarrow \sigma $ as we mentioned in Section~\ref{sec:two}. The
transformation $\pm a\overline{p}^{\prime }a$ converts an element $p^{\prime
}$ of $I^{\prime }$ back to an element of $I$. To check the validity of (\ref%
{eq:25}) we look into the nontrivial example $[-\frac{1}{2}(e_{1}-\sigma
e_{2}-\tau e_{3}),\frac{1}{2}(e_{1}-\sigma e_{2}-\tau e_{3})]^{\ast }$.
Denote by $p$ the element $\frac{1}{2}(e_{1}-\sigma e_{2}-\tau e_{3})$, then
$p^{\prime }$ will be given by $p^{\prime }=\frac{1}{2}(e_{1}-\tau
e_{2}-\sigma e_{3})$ which is an element of the representation $I^{\prime }$%
. Multiplying $\overline{p}^{\prime }$ by $a$ on the left and right we
obtain
\begin{equation}
-a\overline{p}^{\prime }a=\frac{1}{\sqrt{2}}(e_{2}-e_{3})\frac{1}{2}%
(e_{1}-\tau e_{2}-\sigma e_{3})\frac{1}{\sqrt{2}}(e_{2}-e_{3})=\frac{1}{2}%
(e_{1}-\sigma e_{2}-\tau e_{3})=p  \label{eq:26}
\end{equation}%
which converts the $[-p,p]^{\ast }$ element into the general form $%
[p,-ap^{\prime }a]^{\ast }$.

To show that the group represented above is closed under multiplication, we
verify a non-trivial case. Consider two elements $[p_{1},a\overline{p_{1}}%
^{\prime }a]^{\ast }$ and $[p_{2},-a\overline{p_{2}}^{\prime }a]$. In the
product $[p_{1},a\overline{p_{1}}^{\prime }a]^{\ast }[p_{2},-a\overline{p_{2}%
}^{\prime }a]=[-p_{1}ap_{2}a,\overline{p_{2}}^{\prime }a\overline{p_{1}}%
^{\prime }a]^{\ast }$ if we let $p_{3}=-p_{1}ap_{2}a$ and compute $a%
\overline{p_{3}}^{\prime }a$ we obtain $\overline{p_{2}}^{\prime }a\overline{%
p_{1}}^{\prime }a$, so that the product can be written in the form $[p_{3},a%
\overline{p_{3}}^{\prime }a]^{\ast }$. Other products can be handled in an
anologous fashion. Therefore the set of elements in (\ref{eq:25}) form a
group of order $120$. It can be verified that the transformation $a\overline{%
p}^{\prime }a$ exchanges the following conjugacy classes of $I:$
\begin{equation}
\pm 1\leftrightarrow \pm 1,12_{\pm }\leftrightarrow 12_{\pm }^{\prime
},20_{\pm }\leftrightarrow 20_{\pm },15_{\pm }\leftrightarrow 15_{\pm }.
\label{eq:27}
\end{equation}%
The group elements $[p,-a\overline{p}^{\prime }a]$ form a subgroup of order $%
60$ isomorphic to the group $A_{5}$ of even permutations of five letters
whose elements can be written in terms of its five conjugacy classes as
\begin{equation}
\lbrack 1,1],[15_{+},15_{+}],[20_{+},20_{+}],[12_{+},12_{+}^{\prime
}],[12_{+}^{\prime },12_{+}].  \label{eq:28}
\end{equation}%
It is interesting to note the two different realizations of $A_{5}$ in $%
W(H_{4})$ by comparing (\ref{eq:28}) and (\ref{eq:20}). We can check that
the group element $[-1,1]^{\ast }$ which represents reflection with respect
to the root $1$ leaves the set of elements $[p,-a\overline{p}^{\prime }a]$
of $A_{5}$ invariant under conjugation:
\begin{equation}
\lbrack -1,1]^{\ast }[p,-a\overline{p}^{\prime }a][-1,1]^{\ast }=[ap^{\prime
}a,-\overline{p}]=[q,-a\overline{q}^{\prime }a]  \label{eq:29}
\end{equation}%
This proves that the group elements in (\ref{eq:25}) can be written as a
union of two cosets
\begin{equation}
\lbrack p,-a\overline{p}^{\prime }a],[p,-a\overline{p}^{\prime
}a][-1,1]^{\ast }  \label{eq:30}
\end{equation}%
implying that the group structure is
\begin{equation}
W(A_{4})\approx S_{5}\approx A_{5}\rtimes Z_{2}  \label{eq:31}
\end{equation}%
The roots of $H_{4}$ decompose under the Weyl group $W(A_{4})$ as sets of $%
20,\ 20_{+},\ 20_{-},\ 30_{+}$ and $30_{-}$ elements.

The Dynkin diagram symmetry of $SU(5)$ in Figure~\ref{fig:six} exchanges the
simple roots%
\begin{equation}
\begin{array}{lll}
-1 & \leftrightarrow & \frac{1}{2}(e_{1}-\sigma e_{2}-\tau e_{3}) \\
-e_{1} & \leftrightarrow & \frac{1}{2}(1+e_{1}+e_{2}+e_{3})%
\end{array}
\label{eq:32}
\end{equation}%
which can be obtained by the transformation $\gamma =[b,c]$ with $b=\frac{1}{%
2}(-\tau e_{1}+e_{2}+\sigma e_{3})$ and $c=\frac{1}{2}(\sigma e_{1}-\tau
e_{2}-e_{3})$. Then the Weyl group $W(A_{4})$ can be extended to the full
automorphism group $Aut(A_{4})$ of the root system of $SU(5)$ by adjoining
the generator $[b,c]$ to $W(A_{4})$. The extended group is of order $240$
with the structure $Aut(A_{4})\approx W(A_{4})\rtimes Z_{2}$ where $Z_{2}$
is generated by $\gamma $. We repeat the same argument discussed in the
other sections where the product of the largest element $w_{0}$ and $\gamma $
leads to $w_{0}\gamma =[-1,1]=c$. We can then choose $[-1,1]$ as the coset
representative rather then $[b,c]$ in which case the element $[-1,1]$
commutes with the set of elements in (\ref{eq:25}) and transforms all roots
of $H_{4}$ to their negatives. Extention of $W(A_{4})$ either by $[b,c]$ or
by $[-1,1]$ leads to the set of elements
\begin{equation}
\lbrack p,\mp a\overline{p}^{\prime }a],[p,\pm a\overline{p}^{\prime
}a]^{\ast }.  \label{eq:33}
\end{equation}%
When the elements of $Z_{2}^{\prime }=<c=[-1,1]>$ is taken as the coset
representatives, the group of order $240$ obviously manifests itself as a
direct product $W(A_{4})\times Z_{2}^{\prime }$. We can also prove that $%
W(A_{4})\times Z_{2}^{\prime }$ can be embedded in $H_{4}$ in $60$ different
ways.

The maximality of $Aut(A_{4})$ is proven with a method analogous to that
used at the end of Sec. 3.

\end{document}